\documentclass{amsart}
\usepackage{amssymb}
\usepackage{amsfonts}

\newtheorem{theorem}{Theorem}
\theoremstyle{plain}

\numberwithin{equation}{section}
\input{tcilatex}

\begin{document}
\title[Quantum filtering and optimal estimation]{Continuous non-demolition
observation, quantum filtering and optimal estimation}
\author{V.P.Belavkin}
\address{Moscow Institute of Electronics and Mathematics\\
Moscow 109028, USSR}
\email{vpb@maths.nott.ac.uk}
\urladdr{http://www.maths.nott.ac.uk/personal/vpb/}
\thanks{Author thanks to Matt James for support and hospitality in ANU,
Canberra, where this old paper was revised and prepared for the current
publication }
\date{September 20, 2005}
\dedicatory{}
\thanks{This paper was read in the 1st QCMC conference in Paris, November
1990, and was originally published in the conference Proceedings: \textit{%
Quantum Aspects of Optical Communication,} Lecture notes in Physics \textbf{%
45}, pp 131-145, Springer, Berlin 1991.}

\begin{abstract}
A quantum stochastic model for an open dynamical system (quantum receiver)
and output multi-channel of observation with an additive nonvacuum quantum
noise is given. A Master equation for the moment generating operator of the
corresponding instrument is derived and quantum stochastic filtering
equations both for the Heisenberg operators and the reduced density matrix
of the system under the nondemolition observation are found. Thus the
dynamical problem of quantum filtering is generalized for a noncommutative
output process, and a quantum stochastic model and optimal filtering
equation for the dynamical estimation of an input Markovian process is
found. The results are illustrated on an example of optimal estimation of an
input Gaussian diffusion signal, an unknown gravitational force say in a
quantum optical or Weber's antenna for detection and filtering a
gravitational waves.
\end{abstract}

\maketitle

\textbf{Introduction.} The time evolution of quantum system under a
continuous observation can be obtained in the frame work of quantum
stochastic (QS) calculus of output nondemolition processes, firstly
introduced in \cite{bib:b1} and recently developed in a quite general form
in \cite{bib:b2,bib:5,bib:2,bib:b5,bib:9}. A stochastic posterior Schr\"{o}%
dinger wave equation for an observed spinless particle derived in \cite%
{bib:2} by using the quantum filtering method \cite{bib:9}, provided an
explanation of the quantum Zeno paradox \cite{bib:b7,bib:b15}. In this paper
we give a derivation of the reduced wave equation for a Markovian open
system described by Heisenberg quantum stochastic operators $X\left(
t\right) $ with respect to noncommuting Bose output fields $Y\left( s\right)
,s\in \mathbb{R}_{+}$ which are assumed to be nondemolition in the sense 
\cite{bib:2,bib:b5,bib:9} of the commutativity $[X(t),\;Y(s)]=0$ at each
time $t\geq s$. We shall obtain it by a non--unitary dilation of the
characteristic operator of an instrument for the observable output process,
but in contrast to \cite{bib:b8} we restrict ourselves to the diffusion
observation, i.e. to a continuous nondemolition measurement of a quantum
Brownian motion. This gives the possibility to solve the dynamical problems
of quantum detection and estimation theory \cite{bib:b11} as demonstrated in
an example.

\section{The dynamical model}

\typeout{The dynamical model} 

We are going to describe a dynamical model for continuous in time indirect
nondemolition observation of an arbitrary family $\boldsymbol{Q}=\left(
Q_{1},\ldots ,Q_{n}\right) $ of Hermitian operators $Q_{j}=Q_{j}^{\dagger }$%
, acting in initial Hilbert space $\mathcal{H}_{0}$ of an open quantum
system (antenna), with additive $\delta $--correlated Bose type quantum
error-noises in linear $n$-dimensional output channel of observation $%
\mathrm{e}_{j}(t),\;j=1,\ldots ,n$. We shall describe the output quantum
error-noise $\mathbf{e}(t)=\left( \mathrm{e}_{1},\ldots ,\mathrm{e}%
_{n}\right) (t)$ by the components $\widetilde{\mathrm{v}}_{i}\left(
t\right) =\mathrm{e}_{i}\left( t\right) $ of a quantum stochastic process
which is opposite, or inverse to an \emph{input quantum noise} $\mathrm{v}%
_{\cdot }=\left( \mathrm{v}_{i}\right) $ of the same or higher
dimensionality $m\geq n$. The total \emph{quantum output noise} $\widetilde{%
\mathrm{v}}_{\cdot }=\left( \widetilde{\mathrm{v}}_{i}\right) $ as the
opposite to $\mathrm{v}_{\cdot }$ can be defined as a compatible (commuting
with $\mathrm{v}_{\cdot }$) but maximaly closed (maximally entangled with $%
\mathrm{v}_{\cdot }$) quantum stochastic process $\widetilde{\mathrm{v}}%
_{\cdot }$ which simply coincides with the input $\mathrm{v}_{\cdot }$ if it
is is self-compatible, i.e. is classical (having all commutig components $%
\mathrm{v}_{i}$).

Let us describe the quantum input noise $\mathrm{v}_{\cdot }$ by the
classical white noise components $\mathrm{v}_{i}(t)$ represented by
noncommuting Hermitian operator-valued distributions $\mathrm{v}_{i}=\mathrm{%
v}_{i}{}^{\dagger }$. They are completely determined in a quantum Gaussian
state by the first and second moments 
\begin{equation}
\langle \mathrm{v}_{i}(t)\rangle =0,\;\;\langle \mathrm{v}_{i}(t)\mathrm{v}%
_{k}(t^{\prime })\rangle =\kappa _{ik}\delta (t^{\prime }-t).  \label{eq:b1}
\end{equation}%
Here $\kappa _{ik}$ are complex elements of a Hermitian-positve matrix $%
\kappa =\left[ \kappa _{ik}\right] $ of the same or higher dimensionality $%
m\geq n$, with imaginary part $\func{Im}\kappa $ defining the Bose
commutation relations 
\begin{equation*}
\lbrack \mathrm{v}_{i}\left( t\right) ,\;\mathrm{v}_{k}\left( t^{\prime
}\right) ]=2\mathrm{i}\func{Im}\kappa _{ik}\delta \left( t-t^{\prime
}\right) \mathrm{1}\;,\;2\mathrm{i}\func{Im}\kappa _{ik}=\kappa _{ik}-%
\overline{\kappa }_{ik},
\end{equation*}%
such that complex conjugate components $\overline{\kappa _{ij}}=\kappa _{ji}$
define the intensity covariance matrix $\widetilde{\kappa }=\left[ \kappa
_{ji}\right] $ of the output noise $\widetilde{\mathrm{v}}_{\cdot }=\left( 
\widetilde{\mathrm{v}}_{i}\right) :$ 
\begin{equation*}
\langle \widetilde{\mathrm{v}}_{i}(t)\widetilde{\mathrm{v}}_{k}(t^{\prime
})\rangle =\widetilde{\kappa }_{ik}\delta (t-t^{\prime }),\;\;[\widetilde{%
\mathrm{v}}_{i}\left( t^{\prime }\right) ,\;\widetilde{\mathrm{v}}_{k}\left(
t\right) ]=2\mathrm{i}\func{Im}\widetilde{\kappa }_{ik^{\prime }}\delta
\left( t-t^{\prime }\right) \mathrm{1}
\end{equation*}%
as transposed (or complex conjugate, $\overline{\kappa }=\widetilde{\kappa }$%
) to $\kappa $. Thus all output components $\widetilde{\mathrm{v}}_{i}$
commute with all input components $\mathrm{v}_{k}$, and $\widetilde{\mathrm{v%
}}$ must also be maximally correlated with $\mathrm{v}$ in the sense that
the intensities $\gamma _{jk}$ of real covariances%
\begin{equation*}
\langle \widetilde{\mathrm{v}}_{i}(t)\mathrm{v}_{k}(t^{\prime })\rangle
=\gamma _{ik}\delta (t-t^{\prime })=\gamma _{ki}\delta (t^{\prime
}-t)=\langle \mathrm{v}_{k}(t)\widetilde{\mathrm{v}}_{i}(t^{\prime })\rangle 
\end{equation*}%
are the elements of a symmetric $m\times m$-matrix $\gamma =\left[ \gamma
_{ik}\right] $ as the \emph{geometric mean} $\gamma =\left( \kappa \cdot 
\widetilde{\kappa }\right) ^{1/2}$. The geometric mean with $\widetilde{%
\kappa }$ for an invertible $\kappa $ is defined as a Hermitian-positive
matrix $\gamma $ such that $\widetilde{\kappa }=\gamma \kappa ^{-1}\gamma $.
The matix $\gamma $ is symmetric and invertible, with the inverse $\gamma
^{-1}$ =$\left[ \gamma ^{ik}\right] $ defining $\kappa ^{-1}=\gamma ^{-1}%
\widetilde{\kappa }\gamma ^{-1}$ and as the intensity for the covariances%
\begin{equation*}
\langle \widetilde{\mathrm{v}}^{i}(t)\widetilde{\mathrm{v}}^{k}(t^{\prime
})\rangle =\kappa ^{ik}\delta (t-t^{\prime }),\;\;[\widetilde{\mathrm{v}}%
^{i}\left( t\right) ,\;\widetilde{\mathrm{v}}^{k}\left( t^{\prime }\right)
]=2\mathrm{i}\func{Im}\kappa ^{ik}\delta \left( t-t^{\prime }\right) \mathrm{%
1}
\end{equation*}%
of the \emph{contravariant components} $\widetilde{\mathrm{v}}^{i}\left(
t\right) =\gamma ^{i,\cdot }\widetilde{\mathrm{v}}_{\cdot }\left( t\right) $
for the output noise $\widetilde{\mathrm{v}}_{\cdot }\left( t\right) $. (We
assume that Hermitian matrix $\kappa =\left[ \kappa _{ik}\right] $ is
strictly positive, with the inverse $\kappa ^{-1}=\left[ \kappa ^{ik}\right] 
$ corresponding to a finite temperature of the output quantum noise $%
\widetilde{\mathrm{v}}^{\cdot }=\left( \widetilde{\mathrm{v}}^{i}\right) $).

As usual the operator-valued distributions $\mathrm{v}_{i}\left( t\right) $
can be described as generalized derivatives $\mathrm{v}_{i}\left( t\right) =%
\frac{\mathrm{d}}{\mathrm{d}t}\mathrm{v}_{i}^{t}$ of quantum Wiener
vector-process $\mathrm{v}_{\cdot }^{t}$ represented by selfadjoint quantum
stochastic integrators $\mathrm{v}_{i}^{t}=\mathrm{v}_{j}^{t\dagger }$ with $%
\mathrm{v}_{j}^{0}=0$ and independent increments $\mathrm{dv}_{j}^{t}=%
\mathrm{v}_{j}^{t+\mathrm{d}t}-\mathrm{v}_{j}^{t}$ satisfying the
noncommutative multiplication table $\mathrm{dv}_{i}^{t}\mathrm{dv}%
_{k}^{t}=\kappa _{ik}\mathrm{d}t$. Assuming for simplicity that matrix $%
\kappa $ commutes with the transposed $\widetilde{\kappa }$ one can realize
such quantum Wiener noise with respect to the faithful state given by the
vacuum vector $\delta _{\emptyset }$ in a Fock space $\mathcal{F}$ as $%
\mathrm{v}_{j}^{t}=\mathrm{\check{a}}_{j}^{t}+\mathrm{\check{a}}%
_{j}^{t}{}^{\dagger }\equiv 2\Re \mathrm{\check{a}}_{j}^{t}$, the doubled
Hermitian parts of the linear combinations $\mathrm{\check{a}}_{\cdot
}^{t}=\kappa ^{1/2}\mathrm{a}_{t}^{\cdot }$. Here $\mathrm{a}_{t}^{k}$ are
the canonical annihilation integrators which are adjoint to the creation
operators $\mathrm{a}_{j}^{t}{}^{\dagger }$ in the intervals $[0,t)$ defined
on the symmetrical tensors over the complex vector-functions $\alpha _{\cdot
}(t)=\left[ \alpha _{1},\alpha _{2},\ldots \right] (t)$ with%
\begin{equation*}
\left\langle \alpha _{\cdot }|\alpha _{\cdot }\right\rangle =\sum_{ik}\int 
\overline{\alpha }(t)\alpha (t)\mathrm{d}t\equiv \parallel \alpha \parallel
^{2}<\infty \;
\end{equation*}%
as the symmetric tensor multiplication by the indicator function $1_{[0,t)}$%
. The canonical commutation relations 
\begin{equation}
\lbrack \mathrm{a}(\alpha _{\cdot }^{\ast }),\;\mathrm{a}^{\dagger }(\alpha
_{\cdot }^{\prime })]=\left\langle \alpha _{\cdot }|\alpha _{\cdot }^{\prime
}\right\rangle ,\;[\mathrm{a}(\alpha _{\cdot }),\;\mathrm{a}(\alpha _{\cdot
}^{\prime })]=0  \label{eq:b2}
\end{equation}%
then are realized by the quantum stochastic integrals $\mathrm{a}(\alpha
_{\cdot }^{\ast })=\int \bar{\alpha}_{k}(t)\mathrm{da}_{t}^{k}$, $\mathrm{a}%
^{\dagger }(\alpha _{\cdot })=\mathrm{a}(\alpha _{\cdot }^{\ast })^{\dagger }
$. (We use Einstein notations for the convolution $\bar{\alpha}_{k}\beta
^{k}=\sum \bar{\alpha}_{k}\beta ^{k}$ over the indices $k=1,2,\ldots $ in
contrast to the scalar product notations $\beta \cdot \boldsymbol{\alpha }%
^{\ast }$ for the finite sums $\sum_{j=1}^{n}\beta ^{j}\bar{\alpha}_{j}$,
and omit the identity operator $\mathrm{1}$).

The output vector-process $\boldsymbol{\dot{Y}}(t)=\boldsymbol{Q}%
(t)+I_{0}\otimes \mathbf{e}(t)$, defined by the integrals 
\begin{equation}
Y_{j}(t)=\int_{0}^{t}Q_{j}(r)\mathrm{d}r+I_{0}\otimes \mathrm{e}%
_{j}^{t},\;\;j=1,\ldots ,n  \label{eq:b3}
\end{equation}%
of the Heisenberg operators $Q_{j}(t)=U(t)^{\dagger }\left( Q_{j}\otimes 
\mathrm{1}\right) U(t)$, can be realized for a singular coupling of the
system with the Bose fields $\mathrm{a}_{t}^{k}$ by the output observables $%
\mathrm{e}_{j}^{t}=2\Re \mathrm{\hat{a}}_{j}^{t}$, $\mathrm{\hat{a}}_{\cdot
}^{t}=\widetilde{\kappa }^{1/2}\mathrm{a}_{t}^{\cdot }$ in the interaction
picture 
\begin{equation}
Y_{j}(t)=U(t)^{\dagger }\left( I_{0}\otimes \mathrm{e}_{j}^{t}\right)
U(t)=2\Re B_{j}(t),  \label{eq:b4}
\end{equation}%
where $B_{j}(t)=U(t)^{\dagger }(I\otimes \mathrm{a}_{j}^{t})U(t)$ are the
annihilation output processes, introduced in \cite{bib:b2,bib:5,bib:2,bib:b5}%
, and $I_{0}$ is the identity operator in $\mathcal{H}_{0}$. The unitary
evolution $U(t)$ will be described on the tensor product $\mathcal{H}=%
\mathcal{H}_{0}\otimes \mathcal{F}_{\kappa }$ by a Schr\"{o}dinger-It\^{o}
quantum stochastic equation \cite{bib:12} 
\begin{equation}
\mathrm{d}U(t)+KU(t)\mathrm{d}t=\mathrm{i}\left( \frac{1}{\hbar }\ Q\otimes 
\mathrm{d}f\left( \vartheta _{t}\right) -2\Im (L_{k}^{\dagger }\otimes 
\mathrm{d\check{a}}_{t}^{k})\right) U(t),  \label{eq:b5}
\end{equation}%
in terms of the input integrators $\mathrm{\check{a}}_{t}^{\cdot }=%
\widetilde{\kappa }^{-1/2}\mathrm{a}_{t}^{\cdot }=\gamma ^{-1}\mathrm{\check{%
a}}_{\cdot }^{t}$, where $L_{k},\;\;k=1,2,...$ are the operators in $%
\mathcal{H}_{0}$ with $L_{j}+L_{j}^{\dagger }=Q_{j},\;\;j=1,\ldots ,n$, 
\begin{equation*}
K=\frac{1}{2}\ (\frac{\sigma ^{2}}{\hbar ^{2}}Q^{2}f^{\prime }\left(
\vartheta _{t}\right) ^{2}+L_{i}^{\dagger }\widetilde{\kappa }^{ik}L_{k})+%
\frac{\mathrm{i}}{\hbar }H,
\end{equation*}%
$f^{\prime }(\vartheta )=\frac{\mathrm{d}}{\mathrm{d}\vartheta }\
f(\vartheta )$, $H=H^{\dagger }$ is a Hamiltonian of the system, and $%
Q=Q^{\dagger }$ is an operator in $\mathcal{H}_{0}$ of a generalized
coordinate conjugate to the generalized force $f\left( t\right) =$ $\frac{%
\mathrm{d}}{\mathrm{d}t}f\left( \vartheta _{t}\right) $ depending on an
independent input diffusive signal $\vartheta _{t}$, the random position of
a gravitational source say, with $(\mathrm{d}\vartheta _{t})^{2}=\sigma ^{2}%
\mathrm{d}t$. Note that in the case $L_{k}=L_{k}^{\dagger }$ this equation
can be written as%
\begin{equation*}
\mathrm{d}U(t)+KU(t)\mathrm{d}t=\frac{\mathrm{i}}{\hbar }\left( \ Q\otimes 
\mathrm{d}f\left( \vartheta _{t}\right) +Q_{k}\otimes \mathrm{df}%
_{t}^{k}\right) U(t)
\end{equation*}%
in terms of the integrators $\mathrm{f}_{t}^{k}=-\hbar \Im \mathrm{\check{a}}%
_{t}^{k}$ of quantum Langevin forces $\mathrm{f}^{k}\left( t\right) =\frac{%
\mathrm{d}}{\mathrm{d}t}\mathrm{f}_{t}^{k}$ satisfying the canonical
commutation relations%
\begin{equation*}
\left[ \mathrm{f}^{i}\left( t\right) ,\mathrm{e}_{j}\left( t^{\prime
}\right) \right] =\frac{\hbar }{\mathrm{i}}\delta _{j}^{i}\delta \left(
t-t^{\prime }\right) ,\;\;\;[\mathrm{f}^{i}\left( t\right) ,\;\mathrm{f}%
^{k}\left( t^{\prime }\right) ]=\mathrm{i}\frac{\hbar ^{2}}{2}\func{Im}%
\widetilde{\kappa }^{ik}\delta \left( t-t^{\prime }\right) 
\end{equation*}%
corresponding to noncommutative multiplication tables%
\begin{equation*}
\mathrm{de}_{j}^{t}\mathrm{df}_{t}^{k}=\mathrm{i}\hbar \delta _{j}^{k}%
\mathrm{d}t,\mathrm{df}_{t}^{i}\mathrm{df}_{t}^{k}=\left( \hbar /2\right)
^{2}\widetilde{\kappa }^{ik}\mathrm{d}t,
\end{equation*}%
including $\widetilde{\kappa }^{i,0}=\left( 2\sigma f^{\prime }/\hbar
\right) ^{2}\delta _{0}^{i}$ for $\mathrm{f}_{t}^{0}=f\left( \vartheta
_{t}\right) $.

The solution $t\mapsto U(t)$ of the equation (\ref{eq:b4}) for $U(0)=I$ is
adaptive $U(s)=U^{s}\otimes I_{[s}$ with respect to the tensor decomposition 
$\mathcal{H}=\mathcal{H}_{s}\otimes \mathcal{F}_{[s}$, where $\mathcal{H}%
_{s}=\mathcal{H}_{0}\otimes \mathcal{F}_{s}$ and $\mathcal{F}_{s},\;\mathcal{%
F}_{[s}$ are the Fock spaces, generated by the vector--function with $\alpha
_{s}(t)=0,\;\;\forall t\geq s$ and $\alpha _{\lbrack s}(t)=0,\;\;t<s$
respectively. Hence $(I_{0}\otimes \mathrm{e}_{j}^{s})U(t)=U(t)Y_{j}(s)$ for
any $s\leq t$, and 
\begin{eqnarray}
\lbrack Y_{j}(s),\;Y_{k}(t)] &=&U(t)^{\dagger }(I_{0}\otimes \lbrack 2\Re 
\mathrm{\hat{a}}_{i}^{s},\;2\Re \mathrm{\hat{a}}_{k}^{t}])U(t)=2\mathrm{i}s%
\func{Im}\widetilde{\kappa }_{jk}I  \notag \\
\lbrack Y_{j}(s),\;X(t)] &=&U(t)^{\dagger }[I_{0}\otimes \mathrm{e}%
_{j}^{s},\;X\otimes \mathrm{1}]U(t)=0,\;\;\forall s\leq t  \label{eq:b6}
\end{eqnarray}%
for any operator $X(t)=U(t)^{\dagger }(X\otimes \mathrm{1})U(t)$ of the
system in the Heisenberg picture. This proves the nondemolition property of
any observable process $Y_{j}(t)$ with respect to the system. It was
introduced as the nondemolition causality principle for the output quantum
processes in \cite{bib:b1,bib:b2}. Using the quantum It\^{o} formula (\ref%
{eq:b4}) for \cite{bib:2} with $\mathrm{de}_{j}=2\Re \mathrm{d\hat{a}}_{j}$
and multiplication table 
\begin{equation}
\mathrm{d\check{a}}_{t}^{i}\mathrm{d\check{a}}_{t}^{k\dagger }=\widetilde{%
\kappa }^{ik}\mathrm{d}t,\;\;\mathrm{d\check{a}}_{t}^{i}\mathrm{d\hat{a}}%
_{k}^{t\dagger }=\delta _{k}^{i}\mathrm{d}t,\;\;\mathrm{d\hat{a}}_{i}^{t}%
\mathrm{d\hat{a}}_{k}^{t\dagger }=\widetilde{\kappa }_{ik}\mathrm{d}t
\label{eq:b7}
\end{equation}%
with all other products being zero, one can derive the equation (\ref{eq:b3}%
): $\mathrm{d}\boldsymbol{Y}(t)=\boldsymbol{Q}(t)\mathrm{d}t+I_{0}\otimes 
\mathrm{d}\mathbf{e}(t)$, where $\boldsymbol{Q}(t)=[Q_{1},\ldots ,Q_{n}](t)$%
, $Q_{j}(t)=U(t)^{\dagger }(L_{j}+L_{j}^{\dagger })U(t)$.

\begin{theorem}
Let us suppose that the input real--valued signal $\vartheta _{t}$ satisfies
the stochastic differential equation 
\begin{equation}
\mathrm{d}\vartheta _{t}+\upsilon (\vartheta _{t})\mathrm{d}t=\sigma \mathrm{%
dw}_{t},  \label{eq:b8}
\end{equation}%
where $\mathrm{w}_{t}$ is an independent standard Wiener process, defined by
the moments: $\langle \mathrm{w}_{t}\rangle =0,\;\;\langle \mathrm{w}_{s}%
\mathrm{w}_{t}\rangle =s$ for any $s\leq t$ with respect to the vacuum state
as $\mathrm{w}_{t}=2\Re \mathrm{a}_{t}^{0}$ for the canonical annihilation
integrator $\mathrm{a}_{t}^{0}$ in the Fock space $\mathcal{F}$. Then any
twice differentiable in the strong operator topology function $X:\vartheta
\in \mathbf{R}\mapsto X(\vartheta )\in \mathcal{B}(\mathcal{H}_{0})$ in the
Heisenberg picture $X(t)=X^{t}(\vartheta _{t})$, where $X^{t}=U\left(
t\right) ^{\dagger }XU\left( t\right) $, satisfies the following quantum
stochastic Langevin equation 
\begin{eqnarray}
&&\mathrm{d}X(t)+\left( \upsilon \delta X(t)+\frac{\mathrm{i}}{\hbar }\
[X\left( t\right) ,H\left( t\right) ]-\frac{1}{2}\left( \sigma ^{2}\delta
^{2}X(t)+\Lambda _{1}^{\ast t}\left[ X\left( t\right) \right] \right)
\right) \mathrm{d}t  \notag \\
&=&[X(t),\;2\Re L_{k}^{t}]\otimes \frac{\mathrm{i}}{\hbar }\mathrm{df}%
_{t}^{k}+[X(t),\;\mathrm{i}\Im L_{k}^{t}]\otimes \mathrm{dv}_{t}^{k}+\sigma
\delta X(t)\otimes \mathrm{dw}_{t}.  \label{eq:b9}
\end{eqnarray}%
Here $\mathrm{v}_{t}^{i}=2\Re \mathrm{\check{a}}_{t}^{i}=\gamma ^{ik}\mathrm{%
v}_{k}^{t},$ $\delta X(\vartheta )=X^{\prime }(\vartheta )+f^{\prime }\left(
\vartheta \right) [X(\vartheta ),\frac{\mathrm{i}}{\hbar }Q]$, $X^{\prime
}(\vartheta )=\frac{\mathrm{d}}{\mathrm{d}\vartheta }\ X(\vartheta )$ and 
\begin{equation}
\Lambda _{1}^{\ast t}[X^{t}(\vartheta )]=\frac{1}{2}\ \sum_{i,k\geq
1}^{m}\kappa ^{ki}(L_{i}^{t\dagger }[X^{t}(\vartheta
),\;L_{k}^{t}]+[L_{i}^{t\dagger },\;X^{t}(\vartheta )]L_{k}^{t}).
\label{eq:b10}
\end{equation}
\end{theorem}

\section{The reduced evolution}

\typeout{The reduced evolution} 

Let $\mathcal{A}$ denote the input-system algebra which is assumed to be the
von Neumann algebra of all essentially bounded operator-valued functions $%
X:\;\vartheta \in \mathbf{R}\mapsto X(\vartheta )\in \mathcal{B}(\mathcal{H}%
_{0})$, $\mathfrak{\hat{b}}_{t}\subseteq \mathcal{B}\left( \mathcal{F}%
_{t}\right) $ be von Neumann subalgebra generated by the error-noises $%
\left\{ \mathrm{e}_{1}^{s},\ldots ,\mathrm{e}_{n}^{s}\right\} $ for all $%
s\in \lbrack 0,t)$, and $\mathcal{G}_{t}\subseteq \mathcal{F}_{t}$ be
subspace generated by $\mathfrak{\hat{b}}_{t}$ on the vacuum $\delta
_{\emptyset }\in \mathcal{F}_{t}$ for each $t>0$, where $\mathcal{F}_{t}$
are Fock subspaces generated on $\delta _{\emptyset }$ by all input
processes $\mathrm{v}_{s}^{i}$, \textrm{w}$_{s}$ or, equivalently, by forces 
$\left( \mathrm{f}_{s}^{0},\mathrm{f}_{s}^{1},\ldots \right) $ up to time $t$%
. The increasing family $\left\{ \mathfrak{\hat{b}}_{t}:t>0\right\} $ with
respect to the output states induced by $\psi \left( t\right) =U\left(
t\right) \left( \psi _{0}\otimes \delta _{\emptyset }\right) \in \mathcal{F}%
_{t}$, $\psi _{0}\in \mathcal{H}_{0}$ will be called the output filtration
in the Schr\"{o}dinger picture; It is equivalent to the Heisenberg
filtration $\left\{ \mathcal{B}_{t},t>0\right\} $ with $\mathcal{B}%
_{t}\subseteq \mathcal{A}\otimes \mathcal{B}\left( \mathcal{F}_{t}\right) $
generated by the output family $\left\{ Y_{1}\left( s\right) ,\ldots
,Y_{n}\left( s\right) \right\} $ for all $s\in \lbrack 0,t)$ with respect to
the initial states induced by $\psi \left( 0\right) =\psi _{0}\otimes \delta
_{\emptyset }$, since $U_{t}^{\dagger }Y_{j}\left( s\right)
U_{t}=I_{0}\otimes \mathrm{e}_{j}^{s}$ with $U_{t}^{\dagger }=U\left(
t\right) |\mathcal{H}_{t}$ for any $s\leq t$. The \emph{filtered dynamics}
of a quantum stochastic system, described in the Heisenberg picture by
homomorphisms $X\left( t\right) =U_{t}XU_{t}^{\dagger }$ of $\mathcal{A}\ni
X $ into $\mathcal{A}\otimes \mathcal{B}\left( \mathcal{F}_{t}\right) $, is
defined by the cocycle of CP maps $\check{\Phi}_{t}:\mathcal{A}\rightarrow 
\mathcal{A}\otimes \check{\ell}_{t}$, where $\check{\ell}_{t}$ are
(unbounded) commutants of $\mathfrak{\hat{b}}_{t}$ on $\mathcal{G}_{t}$,
such that these dynamics induce the same input-output states on $\mathcal{A}%
\otimes \mathfrak{\hat{b}}_{t}$ with respect to the initial vacuum state:%
\begin{equation}
\epsilon _{\emptyset }\left[ \check{\Phi}_{t}\left[ X\right] \left(
I_{0}\otimes \hat{b}\right) \right] =\epsilon _{\emptyset }\left[
U_{t}\left( X\left( \vartheta _{t}\right) \otimes \hat{b}\right)
U_{t}^{\dagger }\right]  \label{eq:b11}
\end{equation}%
for all $X\in \mathcal{A}$ and $\hat{b}\in \mathfrak{\hat{b}}_{t}$. Here $%
\epsilon _{\emptyset }\left[ \cdot \right] =\left( I_{0}\otimes \delta
_{\emptyset }^{\ast }\right) \left[ \cdot \right] \left( I_{0}\otimes \delta
_{\emptyset }\right) $ is the vacuum (conditional) expectation $\mathcal{A}%
\otimes \mathcal{B}\left( \mathcal{F}\right) \rightarrow \mathcal{A}$ such
that the composition $\rho \circ \epsilon _{\emptyset }\equiv \epsilon
_{\emptyset }\circ \rho $ with any normal state $\rho $ on $\mathcal{A}$ is
the product state $\rho \otimes \epsilon _{\emptyset }$.

Since each $\check{\Phi}_{t}$ is normalized as $\check{\Phi}_{t}\left(
I_{0}\right) =\check{P}_{t}$ to a positive element $\check{P}_{t}\in 
\mathcal{A}\otimes $ $\check{\ell}_{t}$ defining typically unbounded density
operator $\check{p}_{t}=\rho \left[ \check{P}_{t}\right] \in \check{\ell}$
for the output state%
\begin{equation*}
\varsigma _{t}\left( \hat{b}\right) =\rho \left( \epsilon _{\emptyset }\left[
U_{t}\left( I_{0}\otimes \hat{b}\right) U_{t}^{\dagger }\right] \right)
=\epsilon _{\emptyset }\left[ \hat{b}\check{p}_{t}\right]
\end{equation*}%
with respect to the input state $\epsilon _{\emptyset }\left[ \hat{b}\right] 
$ on the algebra $\mathfrak{\hat{b}}_{t}$, it is necessary to give a more
precise meaning of the unbounded commutant $\check{\ell}_{t}$ of $\mathfrak{%
\hat{b}}_{t}$. Due to the nondegeneracy of the covariance matrix-function $%
\left[ \widetilde{\kappa }_{jj^{\prime }}\min \left\{ s,s^{\prime }\right\} %
\right] $ of $\mathbf{e}^{s}:s<t$ for each $t>0$, the cyclic representation $%
\mathfrak{\hat{b}}_{t}|\mathcal{G}_{t}$ on the minimal invariant subspace $%
\mathcal{G}_{t}$ containing $\delta _{\emptyset }$ is faithful in the sense
that $\hat{b}=0$ in $\mathfrak{\hat{b}}_{t}$ if $\hat{b}\delta _{\emptyset
}=0$ in $\mathcal{F}_{t}$. Therefore $\mathfrak{\hat{b}}_{t}$ is transposed
to its bounded commutant $\mathfrak{\hat{b}}_{t}^{\prime }$, coinciding with 
$\mathfrak{\check{b}}_{t}=J\mathfrak{\hat{b}}_{t}J$, where $J$ is an
isometric involution defining the transposition $\mathfrak{\hat{b}}%
_{t}\rightarrow \mathfrak{\check{b}}_{t}$ by $\hat{b}_{t}^{\prime }=J\check{b%
}_{t}^{\dagger }J$ common for all subspaces $\mathcal{G}_{t}$. Moreover, the
von Neumann algebras $\mathfrak{\hat{b}}_{t}$ and $\mathfrak{\hat{b}}%
_{t}^{\prime }$ are in one-to-one correspondence with the achieved Tomita
algebras $\mathfrak{b}_{t}=\mathfrak{\check{b}}_{t}\delta _{\emptyset }$, $%
\mathfrak{b}_{t}^{\prime }=\mathfrak{\hat{b}}_{t}\delta _{\emptyset }$ \cite%
{bib:b12} as dense subspaces of $b=\hat{b}^{\prime }\delta _{\emptyset }$, $%
b^{\prime }=\hat{b}\delta _{\emptyset }$ in $\mathcal{G}_{t}$, where $\check{%
b}=J\hat{b}^{\dagger }J$, $\hat{b}\in \mathfrak{\hat{b}}_{t}$, with common
identity $1:=\delta _{\emptyset }$, involutions $b^{\sharp }:=\check{b}%
^{\dagger }\delta _{\emptyset }$, $b^{\prime \flat }:=\hat{b}^{\dagger
}\delta _{\emptyset }$ such that $b^{\sharp \prime }=b^{\prime \flat }$ and
the norm $\parallel b\parallel _{\infty }:=\left\Vert \check{b}\right\Vert
=\left\Vert \hat{b}\right\Vert \equiv \parallel b^{\prime }\parallel
_{\infty }$.

Let us define a dual space $\ell _{t}$ to the Banach algebra $\mathfrak{b}%
_{t}^{\prime }$ as the completion of $\mathfrak{b}_{t}\subseteq \mathcal{G}%
_{t}$ with respect to the dual norm 
\begin{equation*}
\parallel a\parallel _{1}=\sup \{\left\vert \left\langle b^{\prime
},a\right\rangle _{\emptyset }\right\vert :b\in \mathfrak{b}_{t}^{\prime
},\parallel b^{\prime }\parallel _{\infty }\leq 1\}\equiv \left\Vert \check{a%
}\right\Vert _{\ast },
\end{equation*}%
where the bilinear form is the standard pairing of $\mathfrak{b}_{t}$ and $%
\mathfrak{b}_{t}^{\prime }$, 
\begin{equation}
\left\langle b^{\prime },a\right\rangle :=\epsilon _{\emptyset }\left[ \hat{b%
}\check{a}\right] \equiv \left\langle \hat{b},\check{a}\right\rangle
_{\emptyset },  \label{eq:b12}
\end{equation}%
which we will extend by continuity on all $a\in \ell _{t}\supseteq \mathfrak{%
b}_{t}$. Thus the co-algebra $\check{\ell}_{t}$ of the operator algebra $%
\mathfrak{\hat{b}}_{t}$ is defined as the Banach space of operators $\check{a%
}:b^{\prime }\mapsto \hat{b}a$, mapping $\mathfrak{b}_{t}^{\prime }\subseteq 
\mathcal{G}_{t}$ into $\ell _{t}\supseteq \mathcal{G}_{t}$, bounded with
respect to the dual norm $\left\Vert \cdot \right\Vert _{\ast }$. This space
is a linear span of positive cone $\check{\ell}_{t}=\left\{ \check{p}\geq
0\right\} $ defined by $\left\langle \hat{b}^{\dagger }\hat{b},\check{p}%
\right\rangle _{\emptyset }\geq 0$ for all $\hat{b}\geq 0$, and therefore is
invariant under the right and left action $\check{p}\mapsto \check{b}%
^{\dagger }\check{p}\check{b}$ of $\mathfrak{\check{b}}_{t}$ which is
defined as the dual to the selfaction $\hat{q}\mapsto \hat{b}^{\dagger }\hat{%
q}\hat{b}$ on $\mathfrak{\hat{b}}_{t}$,%
\begin{equation}
\left\langle \hat{q},\check{b}^{\dagger }\hat{p}\check{b}\right\rangle
_{\emptyset }=\left\langle \hat{b}^{\dagger }\hat{q}\hat{b},\hat{p}%
\right\rangle _{\emptyset }\;\;\;\;\forall \hat{q}\in \mathfrak{\hat{b}}_{t},%
\check{p}\in \mathfrak{\check{b}}_{t}  \label{eq:b15}
\end{equation}%
extending the transposed selfaction on $\mathfrak{\check{b}}_{t}$ by $\check{%
b}=J\hat{b}^{\dagger }J$ for all $\hat{b}\in \mathfrak{\hat{b}}_{t}$. The
coalgebra $\check{\ell}_{t}$ is also equipped with involution $\check{a}%
\mapsto \check{a}^{\dagger }$ defined by $\check{a}^{\dagger }b^{\prime }=%
\hat{b}a^{\sharp },\forall \check{a}\in \check{\ell}_{t}$ such that $\langle
b^{\prime },\;a^{\sharp }\rangle =\langle \hat{b},\;\check{a}^{\dagger
}\rangle _{\emptyset }$, and with the identity $\check{1}=\hat{1}$,
corresponding to the vacuum state $\epsilon _{\emptyset }\left( \hat{b}%
\right) =\langle \hat{b},\;\check{1}\rangle _{\emptyset }$. Note that the
elements $\check{a}\in \check{\ell}$ obviously commute with $\hat{b}\in 
\mathfrak{\hat{b}}_{t}$:%
\begin{equation*}
\check{a}\hat{b}c^{\prime }=\hat{b}\hat{c}a=\hat{b}\check{a}c^{\prime
}\;\;\;\;\forall \check{a}\in \ell _{t},\hat{b},\hat{c}\in \mathfrak{\hat{b}}%
_{t},
\end{equation*}%
but are unbounded in the Hilbert space $\mathcal{G}_{t}$ if $a\notin 
\mathfrak{b}_{t}$. However they are densely defined as the bounded kernels
in the Gelfand triple $\mathfrak{b}_{t}\subseteq \mathcal{G}_{t}\subseteq
\ell _{t}$, and $\left\Vert \check{p}\right\Vert _{\ast }=$ $\left\langle 
\hat{1},\check{p}\right\rangle _{\emptyset }$ for any positive element $%
\check{p}\in \check{\ell}_{t}$ which means that it is density operator of a
normal state on $\mathfrak{\hat{b}}_{t}$ if $\left\langle \hat{1},\check{p}%
\right\rangle _{\emptyset }=1$. Moreover, every normal state is uniquely
given by such density, i.e. that the space $\check{\ell}_{t}$ is predual to $%
\mathfrak{\hat{b}}_{t}$, and is preadjoint to $\mathfrak{\check{b}}_{t}$
which we denote as $\check{\ell}_{t}^{\intercal }=\mathfrak{\hat{b}}_{t}$, $%
\check{\ell}_{t}^{\ast }=\mathfrak{\hat{b}}_{t}$. In most cases the density
operator $\check{p}\in \check{\ell}$ of an output state $\varsigma
_{t}\left( \hat{b}\right) =\left\langle \hat{b},\check{p}\right\rangle
_{\emptyset }$ has the range $\check{p}\ell $ in $\mathcal{F}_{t}$, as it is
in the case of a majorized positive form $\varsigma (\hat{b}^{\dagger }\hat{b%
})\leq \lambda \epsilon _{\emptyset }(\hat{b}^{\dagger }\hat{b})$ for a $%
\lambda >0$, corresponding to the bounded $\check{p}$ on $\mathcal{F}_{t}$:$%
\;\left\Vert \check{p}\right\Vert \leq \lambda $; moreover, any operator $%
\check{p}\in \check{\ell}$ can be approximated by the density operators from
the bounded commutant $\mathfrak{\check{b}}_{t}$ in the norm $\left\Vert 
\check{p}\right\Vert _{\ast }=\parallel p\parallel _{1}$, where $p=\check{p}%
\delta _{\emptyset }$.

In order to derive a quantum stochastic equation for the reduced dynamics $%
\check{\Phi}_{t}$, let us find the differential evolution for the factorial
generating map $\Phi _{t}^{(\beta )}:\;\mathcal{A}\rightarrow \mathcal{A}$, 
\begin{equation}
\Phi _{t}^{(\beta )}[X]=\epsilon _{\emptyset }\left[ \check{\Phi}%
_{t}[X](I\otimes \hat{z}_{t}^{(\beta )})\right] =\langle \delta _{\emptyset
}|X^{(\beta )}(t)|\delta _{\emptyset }\rangle .  \label{eq:b13}
\end{equation}%
Here $\hat{b}_{t}=\hat{z}_{t}^{(\beta )}$ are exponential elements, defined
by the Wick (normal) exponent 
\begin{equation}
\hat{z}_{t}^{(\beta )}=e^{\mathrm{\hat{a}}^{\dagger }(\beta _{t})}e^{\mathrm{%
\hat{a}}(\beta _{t})}\equiv :\exp \left[ \mathrm{e}_{t}(\beta )\right] :
\label{eq:b14}
\end{equation}%
of the observable $\mathrm{e}(\beta _{t})=\mathrm{\hat{a}}\left( \beta
_{t}\right) +\mathrm{\hat{a}}^{\dagger }\left( \gamma \beta _{t}\right)
\equiv \int_{0}^{t}\beta (r)\cdot $\textrm{d}$\mathbf{e}^{r}$, where $\beta
=\left( \beta ^{i}\right) $ is a column of locally square-integrable
functions with $\beta _{s}^{j}(t)=0$ for $t\geq s,j>n$, $\beta
_{s}^{j}(r)=\beta ^{j}(r)$ for $r<s$, and $X^{(\beta )}(t)=U_{t}(X(\vartheta
_{t})\otimes \hat{z}_{t}^{(\beta )})U_{t}^{\dagger }=X(t)Z(t)$. Taking into
account the equation (\ref{eq:b9}) and 
\begin{equation*}
\mathrm{d}Z(t)=Z(t)\beta ^{j}(t)(I\otimes 2\mathrm{d}\Re \mathrm{\hat{a}}%
_{j}^{t}+U_{t}Q_{j}U_{t}^{\dagger }\mathrm{d}t)
\end{equation*}%
for $Z(t)=U_{t}\left( I_{0}\otimes \hat{z}_{t}^{(\beta )}\right)
U_{t}^{\dagger }$, one can obtain by quantum It\^{o} formula \cite{bib:12} 
\begin{eqnarray*}
\mathrm{d}(X(t)Z(t)) &=&U_{t}(\Lambda ^{\ast }[X](\vartheta _{t})+(%
\boldsymbol{L}^{\dagger }X_{t}^{(\beta )}+X_{t}^{(\beta )}\boldsymbol{L}%
)\beta (t))U_{t}^{\dagger }\mathrm{d}t \\
&&+\sigma U_{t}\delta X_{t}^{(\beta )}U_{t}^{\dagger }\otimes \mathrm{dw}%
_{t}+U_{t}X_{t}^{\left( \beta \right) }U_{t}^{\dagger }\beta ^{j}(t)\otimes 
\mathrm{d}2\Re \mathrm{\hat{a}}_{j}^{t} \\
&&+U_{t}[L_{k}^{\dagger },X_{t}^{(\beta )}]U_{t}^{\dagger }\otimes \mathrm{d%
\check{a}}_{t}^{k}+\mathrm{d\check{a}}_{t}^{k}{}^{\dagger }\otimes
U_{t}[X_{t}^{(\beta )},\;L_{k}]U_{t}^{\dagger },
\end{eqnarray*}%
where $X_{t}^{(\beta )}=X(\vartheta _{t})\otimes \hat{z}_{t}^{(\beta
)},\;\;\delta =\partial /\partial \vartheta _{t}+f^{\prime }\ [\cdot ,\frac{%
\mathrm{i}}{\hbar }Q]$. Hence the map $\Phi _{t}^{(\beta )}:\;X\in \mathcal{A%
}\mapsto \langle \delta _{\emptyset }|U_{t}X_{t}^{(\beta )}U_{t}^{\dagger
}|\delta _{\emptyset }\rangle $ satisfies the equation 
\begin{equation}
\frac{\mathrm{d}}{\mathrm{d}t}\ \Phi _{t}^{(\beta )}[X]=\Phi _{t}^{(\beta
)}[\Lambda ^{\ast }[X]+(\boldsymbol{L}^{\dagger }X+X\boldsymbol{L})%
\boldsymbol{\beta }(t)],\;\;\;\Phi _{0}^{(\beta )}[X]=X,  \label{eq:b16}
\end{equation}%
where $\Lambda ^{\ast }\left[ X\right] \left( \vartheta \right) =-\upsilon
\left( \vartheta \right) \delta X\left( \vartheta \right) -\frac{\mathrm{i}}{%
\hbar }\ [X\left( \vartheta \right) ,H]+\frac{1}{2}\left( \sigma ^{2}\
\delta ^{2}X\left( \vartheta \right) +\Lambda _{1}^{\ast }\left[ X\left(
\vartheta \right) \right] \right) $.

In the same way, using the It\^{o} formula for the product $\check{G}_{t}%
\hat{Z}_{t}$ of $\check{G}_{t}=\check{\Phi}_{t}[X]$ and the Wick exponent (%
\ref{eq:b14}), one can obtain the equation for (\ref{eq:b13}) if 
\begin{equation}
\mathrm{d}\check{\Phi}_{t}[X]-\check{\Phi}_{t}[\Lambda ^{\ast }[X]]\mathrm{d}%
t=\sum_{j=1}^{n}\check{\Phi}_{t}[L_{j}^{\dagger }X+XL_{j}]\otimes \mathrm{dv}%
_{t}^{j},  \label{eq:b17}
\end{equation}%
where the operators $\mathrm{v}_{t}^{j}=2\Re \mathrm{\check{a}}_{t}^{j}$,
satisfying the canonical commutation relations 
\begin{equation*}
\lbrack \mathrm{v}_{t}^{j^{\prime }},\;\mathrm{v}_{t^{\prime }}^{j}]=2%
\mathrm{i\func{Im}}\widetilde{\kappa }^{j^{\prime }j}\min \{t,\;t^{\prime
}\},\;\;\left[ \mathrm{v}_{t^{\prime }}^{j},\mathrm{e}_{j^{\prime }}^{t}%
\right] =0,
\end{equation*}%
generate the predual coalgebra $\check{\ell}_{t}$ as the unbounded commutant
of $\mathfrak{\hat{b}}_{t}=\{\mathrm{e}_{j}^{s}:s<t\}^{\prime \prime }$
since $\mathrm{v}_{s}^{j},j=1,\ldots ,n$ leave the subspaces $\mathcal{G}%
_{t} $ invariant for any $t\geq s$, and $\{\mathrm{e}_{j}^{s}:s<t,j\leq
n\}^{\prime }=\mathfrak{\check{b}}_{t}=\{\mathrm{v}_{s}^{j}:s<t,j\leq
n\}^{\prime \prime }$ on $\mathcal{G}_{t}$. This proves the following
theorem.

\begin{theorem}
Let the initial state $\rho :\mathcal{A}\rightarrow \mathbf{C}$ be a normal
one, described by a density $\varrho :\;\vartheta \mapsto \varrho (\vartheta
)$ with values in trace--class operators on $\mathcal{H}_{0}$ such that$\rho
\left( X\right) =$ $\int \mathrm{Tr}\varrho (\vartheta )X\left( \vartheta
\right) $\textrm{d}$\vartheta $. Then the conditional state 
\begin{equation}
\check{\phi}_{t}[X]=\int \mathrm{Tr}_{\mathcal{H}_{0}}[\varrho (\vartheta )%
\check{\Phi}_{t}[X](\vartheta )]\mathrm{d}\vartheta  \label{eq:b18}
\end{equation}%
is described in the standard representation by the operator-function $\check{%
\varphi}_{t}(\vartheta )$ as $\check{\phi}_{t}[X]=\int \mathrm{Tr}_{\mathcal{%
H}_{0}}[\check{\varphi}_{t}(\vartheta )X(\vartheta )]$\textrm{d}$\vartheta $
satisfying the quantum stochastic equation 
\begin{equation}
\mathrm{d}\check{\varphi}_{t}(\vartheta )=\Lambda \lbrack \check{\varphi}%
_{t}](\vartheta )\mathrm{d}t+\sum_{j=1}^{n}(L_{j}\check{\varphi}%
_{t}(\vartheta )+\check{\varphi}_{t}(\vartheta )L_{j}^{\dagger })\otimes 
\mathrm{dv}_{t}^{j}.  \label{eq:b19}
\end{equation}%
with $\check{\varphi}_{0}(\vartheta )=\varrho (\vartheta )$. Here the
quantum stochastic differentials $\mathrm{dv}_{t}^{j}$ together with $%
\mathrm{de}_{j}^{t}$ have the canonical multiplication table%
\begin{equation*}
\mathrm{dv}_{t}^{j}\mathrm{dv}_{t}^{j^{\prime }}=\widetilde{\kappa }%
^{jj^{\prime }}\mathrm{d}t,\;\;\mathrm{dv}_{t}^{j}\mathrm{de}_{j^{\prime
}}^{t}=\delta _{j^{\prime }}^{j}\mathrm{d}t,\;\;\mathrm{de}_{j}^{t}\mathrm{de%
}_{j^{\prime }}^{t}=\widetilde{\kappa }_{jj^{\prime }}\mathrm{d}t,
\end{equation*}%
and $\Lambda \left[ \varphi \right] =\delta \left( \upsilon \varphi \right) +%
\frac{\mathrm{i}}{\hbar }\ [\varphi ,H]+\frac{1}{2}\left( \sigma ^{2}\
\delta ^{2}\varphi +\Lambda _{1}\left[ \varphi \right] \right) $ is
preadjoint generator defined on $\mathcal{A}_{\ast }$ by $\delta \varphi
\left( \vartheta \right) =\varphi ^{\prime }\left( \vartheta \right)
+f^{\prime }\left( \vartheta \right) \left[ \varphi \left( \vartheta \right)
,\frac{\mathrm{i}}{\hbar }Q\right] $, 
\begin{eqnarray}
\delta ^{2}\varphi (\vartheta ) &=&\varphi ^{\prime \prime }(\vartheta
)+\left( 2f^{\prime }\left( \vartheta \right) \ +f^{\prime \prime }\left(
\vartheta \right) \right) [\varphi ^{\prime }(\vartheta ),\frac{\mathrm{i}}{%
\hbar }Q]+f^{\prime }\left( \vartheta \right) ^{2}\ [[\varphi (\vartheta ),%
\frac{\mathrm{i}}{\hbar }Q],\frac{\mathrm{i}}{\hbar }Q],  \notag \\
&&\Lambda _{1}\left[ \varphi \right] \left( \vartheta \right) =\sum_{i,k\geq
0}\kappa ^{ik}([L_{i},\;\varphi (\vartheta )L_{k}^{\dagger }]+[L_{i}\varphi
(\vartheta ),\;L_{k}^{\dagger }]),  \label{eq:b20}
\end{eqnarray}%
It is normalized to a positive martingale $\check{p}_{t}=\int \mathrm{Tr}_{%
\mathcal{H}_{0}}\check{\varphi}_{t}(\vartheta )$\textrm{d}$\vartheta \in 
\check{\ell}_{t}$ as the density operator defining by (\ref{eq:b11}) the
output state $\langle \hat{b}(t)\rangle =\rho \lbrack \langle \delta
_{\emptyset }|B\left( t\right) |\delta _{\emptyset }\rangle ]$ on $\ell
^{\prime }\ni z$ for $B(t)=U(t)^{\dagger }(I\otimes \hat{b})U(t)$ with
respect to the vacuum state vector $\delta _{\emptyset }\in \mathcal{F}$
simply as $\langle B(t)\rangle =\langle \check{p}_{t},\hat{b}\rangle
_{\emptyset }$.\newline
\end{theorem}

Note that since all the input components $\mathrm{v}_{t}^{i},i=1,\ldots ,n$
commute with the output components $\mathrm{e}_{j}^{t}$ and have with them
zero correlations and thus are independent of $\mathrm{e}_{j}^{t}$ unless $%
i=j$, they generate the same subspaces $\mathcal{G}_{t}$ as $\mathrm{e}%
_{j}^{t},j=1,\ldots ,n$. On these subspaces they simply coincide with the
transposed $\mathrm{e}_{t}^{i\prime }=J\mathrm{e}_{t}^{i}J=\mathrm{e}%
_{t}^{i\prime \dagger }$ to the contravariant components $\mathrm{e}%
_{t}^{i}=2\Re \mathrm{\hat{a}}_{t}^{i}$ of the output noises $\mathrm{e}%
_{k}^{t}$, having the same autocorrelation and mutual correlation matrices
with $\mathrm{e}_{k}^{t}$ as $\mathrm{v}_{t}^{i}$ and being given by $%
\mathrm{\hat{a}}_{t}^{\cdot }=\kappa ^{-1/2}\mathrm{a}_{t}^{\cdot }=\gamma
^{-1}\mathrm{\hat{a}}_{\cdot }^{t}$ such that $\mathrm{e}_{t}^{i}=%
\sum_{k=1}^{m}\gamma ^{ik}\mathrm{e}_{k}^{t}$. Thus in the filtering
equation (\ref{eq:b19}) the input noises $\mathrm{v}_{t}^{j}$ on $\mathcal{G}%
_{t}$ can be replaced by $\mathrm{e}_{t}^{j\prime }$ which can be given in
the Gaussian case as linear combinations of the transposed noises $\mathrm{e}%
_{j}^{\prime }=J\mathrm{e}_{j}J=\mathrm{e}_{j}^{\prime \dagger }$.

\section{The optimal measurement.}

\typeout{The optimal measurement.} 

The quantum filter defines a quantum measurement on the output of the system
at a time instant $t\in \mathbf{R}_{+}$. In general it is described by a $%
\mathcal{B}_{t}$-valued positive measure $M(t,\mathrm{d}x)$ on a Borel space 
$\mathcal{X}$, normalized to the identity operator in $\mathcal{H}=\mathcal{H%
}_{0}\otimes \mathcal{F}$:$\;\int M(t,\mathrm{d}x)=I$. The problem of
optimal quantum observation is the problem of finding the optimal
measurement $M^{\circ }(t)$, giving the minimal value of the mean 
\begin{equation}
\int \langle S(t,x)M(t,\mathrm{d}x)\rangle =\int \langle m_{t}^{\prime }(%
\mathrm{d}x),c_{t}(x)\rangle =\int \langle \hat{m}_{t}(\mathrm{d}x),\check{c}%
_{t}(x)\rangle _{\emptyset }  \label{eq:b21}
\end{equation}%
for an $\mathcal{A}$-valued measurable function $S:x\in \mathcal{X}\mapsto
S(x)=S(x)^{\dagger }$ in the Heisenberg picture $S(t,x)=U_{t}S(x,\vartheta
_{t})U_{t}^{\dagger }$ with respect to an initial state $\langle \cdot
\rangle =\rho \circ \epsilon _{\emptyset }\left[ \cdot \right] $ on $%
\mathcal{A}\otimes \mathcal{B}_{t}$. Here the mean (\ref{eq:b21}) is given
in the standard representation $\mathcal{B}_{t}\rightarrow \mathfrak{\hat{b}}%
_{t}$ according to (\ref{eq:b11}) as the integral of the pairing (\ref%
{eq:b12}) for 
\begin{equation}
m_{t}^{\prime }(\mathrm{d}x)=\hat{m}_{t}(\mathrm{d}x)\delta _{\emptyset
},\;\;c_{t}(x)=\check{c}_{t}(x)\delta _{\emptyset },  \label{eq:b22}
\end{equation}%
where $\hat{m}_{t}$ defines the measure $M(t)=U_{t}(I\otimes \hat{m}%
_{t})U_{t}^{\dagger }$ in the Schr\"{o}dinger picture, and 
\begin{equation}
\check{c}_{t}(x)=\int \mathrm{Tr}_{\mathcal{H}_{0}}[\check{\varphi}%
_{t}(\vartheta )S(x,\vartheta )]\mathrm{d}\vartheta =\check{\phi}%
_{t}[S_{t}(x)].  \label{eq:b23}
\end{equation}%
The duality principle gives the necessary and sufficient conditions \cite%
{bib:b13} of optimality 
\begin{equation}
\int \langle \hat{m}_{t}^{\circ }(\mathrm{d}x),\check{c}_{t}(x)\rangle
=\inf_{\hat{m}\geq 0}\{\int \langle \hat{m}_{t}(\mathrm{d}x),\check{c}%
_{t}(x)\rangle :\int \hat{m}_{t}^{\circ }(\mathrm{d}x)=\hat{1}\}
\label{eq:b24}
\end{equation}%
of a positive $\mathfrak{\hat{b}}_{t}$-valued measure $\hat{m}_{t}^{\circ
}\geq 0$ with $\int \hat{m}_{t}^{\circ }(\mathrm{d}x)=\hat{1}$ as the
conditions for the dual problem 
\begin{equation*}
\sup \{\langle \hat{1},\check{l}\rangle _{\emptyset }:\check{l}_{t}\leq 
\check{c}_{t}(x),x\in \mathcal{X}\}=\langle \hat{1},\check{l}_{t}^{\circ
}\rangle _{\emptyset },
\end{equation*}%
where $\check{l}_{t}=\check{l}_{t}^{\circ }$ defined by $\check{l}%
_{t}b^{\prime }=\hat{b}l_{t}$ is the operator $\check{l}_{t}^{\circ }\leq 
\check{c}_{t}(x),\;\forall x\in \mathcal{X}$, for which $\int \langle \hat{m}%
_{t}^{\circ }(\mathrm{d}x),\check{c}_{t}(x)\rangle _{\emptyset }=\langle 
\hat{1},\check{l}_{t}^{\circ }\rangle _{\emptyset }$. The last condition of
optimality can be written in the form of the equation 
\begin{equation*}
(\check{c}_{t}(x)-\check{l}_{t})m_{t}^{\prime }(\mathrm{d}x),\;\text{or}\;\;%
\hat{m}_{t}(\mathrm{d}x)(c_{t}(x)-l_{t})=0.
\end{equation*}

Let us consider now the problem of optimal estimation \cite{bib:b11} of a
selfadjoint operator $X(t)=X^{t}(\vartheta _{t})$, given in the Schr\"{o}%
dinger picture by a $\mathcal{B}(\mathcal{H}_{0})$-valued function $%
X(\vartheta )=X(\vartheta )^{\dagger }$, with $x=\lambda \in \mathbf{R}%
,\;S(x,\vartheta )=(X(\vartheta )-xI)^{2}$. One can treat in such a way the
problem of filtering of a real measurable function $x(\vartheta _{t})$ of
the input diffusion signal $\vartheta _{t}$ taking $X(\vartheta
)=x(\vartheta )I$. In order to formulate the optimal estimate in terms of
the measurement of the optimal output observable $\hat{x}_{t}^{\circ }\in 
\mathfrak{\hat{b}}_{t}$ as an appropriate posterior mean of $X\left(
t\right) $ we need the quantum stochastic equation for $\hat{\phi}_{t}\left[
X\right] =J\check{\phi}\left[ X^{\dagger }\right] J$ in terms of the output
noise $\mathrm{e}_{i}^{t}$. It can be obtained by complex conjugation%
\begin{equation}
\mathrm{d}\hat{\phi}_{t}\left[ X\right] =\hat{\phi}_{t}\circ \Lambda ^{\ast
}[X](\vartheta )\mathrm{d}t+\sum_{j=1}^{n}\hat{\phi}_{t}\left[
L_{j}X+XL_{j}^{\dagger }\right] \otimes \mathrm{de}_{t}^{j}  \label{eq:b25}
\end{equation}%
of the equation for the CP map $\check{\phi}_{t}:\mathcal{A}\rightarrow 
\mathfrak{\check{b}}_{t}$ described in (\ref{eq:b18}) by the density
operator $\check{\varphi}_{t}$ satisfying the equation (\ref{eq:b19}). Here $%
\mathrm{e}_{t}^{j}=\sum_{j=1}^{n}\theta ^{jj^{\prime }}\mathrm{e}_{j^{\prime
}}^{t}$, where $\left[ \theta ^{jj^{\prime }}\right] \equiv \boldsymbol{%
\theta }^{-1}$ is real symmetric $n\times n$-matrix with the inverse $%
\boldsymbol{\theta }=\left[ \theta _{jj^{\prime }}\right] $ as an intensity
matrix of the standard covariaces 
\begin{equation*}
\langle \mathrm{e}_{j^{\prime }}(t^{\prime })\mathrm{e}_{j}^{\prime
}(t)\rangle =\theta _{j^{\prime }j}\delta (t^{\prime }-t)=\theta
_{jj^{\prime }}\delta \left( t-t^{\prime }\right) =\left\langle \mathrm{e}%
_{j}^{\prime }\left( t\right) \mathrm{e}_{j^{\prime }}\left( t^{\prime
}\right) \right\rangle
\end{equation*}%
of the output noises $\mathrm{e}_{j^{\prime }}$ with the transposed
components $\mathrm{e}_{j}^{\prime }=J\mathrm{e}_{j}J=\mathrm{e}_{j}^{\prime
\dagger }$ such that $\boldsymbol{\theta \kappa }^{-1}\boldsymbol{\theta }%
^{\intercal }=\overline{\boldsymbol{\kappa }}$ for the $n\times n$-submatrix 
$\boldsymbol{\kappa }$ of $\kappa $. (It is the geometric mean of $%
\boldsymbol{\kappa }$ and $\overline{\boldsymbol{\kappa }}$, e.g. $%
\boldsymbol{\theta }=\boldsymbol{\kappa }^{\frac{1}{2}}\overline{\boldsymbol{%
\kappa }}^{\frac{1}{2}}$ if $\boldsymbol{\kappa }$ and $\overline{%
\boldsymbol{\kappa }}$ commute.) Note that since in general $\boldsymbol{%
\theta }$ is smaller than the $n\times n$-submatrix $\boldsymbol{\gamma }=%
\left[ \gamma _{jj^{\prime }}\right] $ of the mutual covariance matrix $%
\gamma $ for $\mathrm{e}_{i}=\widetilde{\mathrm{v}}_{i}$ and $\mathrm{v}_{k}$%
, $\mathrm{e}_{j}^{\prime }\neq \mathrm{v}_{j}$ unless $m=n$, and similar $%
\mathrm{e}_{t}^{j}\neq \widetilde{\mathrm{v}}_{t}^{j}$ since $\mathrm{e}%
_{t}^{\prime }=\mathrm{v}_{t}$ on $\mathcal{G}_{t}$ and $\boldsymbol{\gamma }%
^{-1}\leq \boldsymbol{\theta }^{-1}$.

\begin{theorem}
The solution of the optimization problem (\ref{eq:b24}) for the quadratic
cost function 
\begin{equation*}
\check{c}_{t}(\lambda )=\lambda ^{2}\check{p}_{t}-2\lambda \check{\phi}%
_{t}[X]+\check{\phi}_{t}[X^{2}]
\end{equation*}%
is given by the spectral measure $\hat{m}_{t}^{\circ }$ of a selfadjoint
operator $\hat{x}_{t}^{\circ }\in \mathfrak{\hat{b}}_{t}$ defined by $\hat{x}%
_{t}^{\circ }=J\check{x}_{t}J$ as transposed to the operator $\check{x}_{t}=%
\check{x}_{t}^{\dagger }$ resolving in $\mathfrak{\check{b}}_{t}$ the
equation 
\begin{equation}
\check{x}_{t}\check{p}_{t}+\check{p}_{t}\check{x}_{t}=2\int \mathrm{Tr}_{%
\mathcal{H}_{0}}[X\left( \vartheta \right) \check{\varphi}_{t}(\vartheta )]%
\mathrm{d}\vartheta .  \label{eq:b26}
\end{equation}

It is given as the symmetric posterior expectation $\hat{x}^{\circ }=\hat{%
\rho}_{t}\left[ X\right] $ by an operator-function $\vartheta \mapsto \hat{%
\varrho}_{t}\left( \vartheta \right) \in \mathcal{B}_{\ast }\left( \mathcal{H%
}\right) \otimes \hat{\ell}_{t}$ with $\hat{\ell}_{t}=J\check{\ell}_{t}J$ as
the density of the solution to the equation%
\begin{equation}
\hat{p}_{t}\hat{\rho}_{t}\left[ X\right] +\hat{\varrho}_{t}\left[ X\right] 
\hat{p}_{t}=2\hat{\phi}_{t}\left[ X\right] \;\;\;\;\forall X\in \mathcal{A},
\label{eq:b27}
\end{equation}%
where $\hat{p}_{t}=J\check{p}_{t}J$. The symmetric posterior density $\hat{%
\varrho}_{t}=\hat{\varrho}_{t}^{\dagger }$ satisfies the nonlinear quantum
stochastic equation 
\begin{equation}
\mathrm{d}\hat{\varrho}_{t}(\vartheta )=\Gamma _{t}[\hat{\varrho}%
_{t}](\vartheta )\mathrm{d}t+\sum_{j=1}^{n}\Xi _{t}^{j}[\hat{\varrho}%
_{t}]\otimes \mathrm{de}_{j}^{t},\;\;\hat{\varrho}_{0}(\vartheta )=\varrho
(\vartheta ),  \label{eq:b28}
\end{equation}%
where $\Gamma _{t}(\vartheta )$ and $\Xi _{t}^{j}(\vartheta )$ are defined
by the solutions to the equations 
\begin{eqnarray}
\Re \hat{P}_{t}\Xi ^{j}[\hat{\varrho}](\vartheta ) &=&\Re \left( \hat{P}%
_{t}(L^{j}\hat{\varrho}(\vartheta )+\hat{\varrho}(\vartheta )L^{j\dagger
})-P_{t}^{j}\varrho (\vartheta )\right) ,  \notag \\
\Re \hat{P}_{t}\Gamma _{t}[\hat{\varrho}](\vartheta ) &=&\Re (\hat{P}%
_{t}\Lambda \lbrack \hat{\varrho}](\vartheta )-\sum_{i,j=1}^{n}\kappa ^{ji}%
\hat{P}_{t}^{i}\Xi _{t}^{j}[\hat{\varrho}]),  \label{eq:b29}
\end{eqnarray}%
with $L^{j}=\sum_{i=1}^{n}L_{i}\theta ^{ij}$, $\hat{P}_{t}=I_{0}\otimes \hat{%
p}_{t}$ and $\hat{P}_{t}^{j}=I_{0}\otimes \hat{p}_{t}^{j}$ defined by $%
Q^{j}=L^{j}+L^{j\dagger }$ and $\hat{\varrho}_{t}$ as 
\begin{equation}
\hat{p}_{t}^{j}=\Re \lbrack \hat{p}_{t}\hat{q}_{t}^{j}],\;\hat{q}%
_{t}^{j}=\int \mathrm{Tr}_{\mathcal{H}_{0}}[\hat{\varrho}_{t}(\vartheta
)Q^{j}]\mathrm{d}\vartheta .  \label{eq:b30}
\end{equation}
\end{theorem}

\textbf{Proof.} Denoting $\hat{u}_{t}=\int \lambda \hat{m}_{t}(\mathrm{d}%
\lambda )$ for an orthogonal projective-valued measure $\hat{m}_{t}(\mathrm{d%
}\lambda )\in \mathfrak{\hat{b}}_{t}$, one obtains 
\begin{eqnarray*}
\int \langle \hat{m}_{t}(\mathrm{d}\lambda ),\check{c}_{t}(\lambda )\rangle
&=&\langle \hat{u}_{t}^{2},\check{p}_{t}\rangle -2\langle \hat{u}_{t},\check{%
\phi}_{t}[X]\rangle +\langle 1,\check{\phi}_{t}[X^{2}]\rangle \\
&=&\langle \hat{u}_{t}^{2},\check{p}_{t}\rangle -2\langle \Re \left( \hat{x}%
_{t}\hat{u}_{t}\right) ,\check{p}_{t}\rangle +\langle 1,\check{\phi}%
_{t}[X^{2}]\rangle \geq \langle \hat{1},\check{l}_{t}^{\circ }\rangle ,
\end{eqnarray*}%
where $\hat{x}_{t}$ is defined by the duality (\ref{eq:b15}) as the
transposed to the solution $\check{x}$ of the equation (\ref{eq:b26})
written as $\Re \check{p}_{t}\check{x}_{t}=\check{\phi}_{t}[X]$, and $\check{%
l}_{t}^{\circ }\in \check{\ell}_{t}$ is defined as%
\begin{equation*}
\check{l}_{t}^{\circ }=\check{\phi}_{t}[X^{2}]-\check{x}_{t}\check{p}_{t}%
\check{x}_{t}.
\end{equation*}%
The inequality is due to positivity of $\check{p}_{t}=\check{\phi}\left[
I_{0}\right] \in \check{\ell}_{t}$ and $(\hat{u}-\hat{x})^{2}\in \mathfrak{%
\hat{b}}_{t}$: 
\begin{equation*}
\int \langle \hat{m}_{t}(\mathrm{d}\lambda ),\check{c}_{t}(\lambda )\rangle
-\langle \hat{1},\check{l}_{t}^{\circ }\rangle =\langle (\hat{u}_{t}-\hat{x}%
_{t})^{2},\;\check{p}_{t}\rangle \geq 0,
\end{equation*}%
and the equality is achieved at the spectral measure $\hat{m}_{t}^{\circ }$
of the selfadjoint operator $\hat{x}_{t}\in \mathfrak{\hat{b}}_{t}$ defining 
$\hat{u}_{t}$ as $\hat{x}_{t}$.

Representing the solution of the equation (\ref{eq:b26}) as%
\begin{equation*}
\check{x}_{t}=\int \mathrm{Tr}_{\mathcal{H}_{0}}X\left( \vartheta \right) 
\check{\varrho}_{t}\left( \vartheta \right) \mathrm{d}\vartheta \equiv 
\check{\rho}_{t}\left[ X\right]
\end{equation*}%
in terms of the density function $\check{\varrho}_{t}\in \mathcal{A}_{\ast
}\otimes \check{\ell}_{t}$ we note that $\check{\rho}_{t}$ satisfies the
equation transposed to (\ref{eq:b27}), and therefore $\hat{\rho}_{t}\left[ X%
\right] =J\check{\rho}_{t}\left[ X^{\dagger }\right] J$ satisfies the
equation (\ref{eq:b27}).

Looking for the operator $\hat{x}_{t}=\hat{\rho}_{t}\left[ X\right] $ as the
solution of a quantum stochastic equation%
\begin{equation*}
\mathrm{d}\hat{x}_{t}=\hat{g}_{t}\mathrm{d}t+\sum_{j=1}^{n}\hat{c}%
_{t}^{j}\otimes \mathrm{de}_{j}^{t}
\end{equation*}%
with some $\hat{g}_{t}=\hat{g}_{t}^{\dagger },\;\hat{c}_{t}^{j}=\hat{c}%
_{t}^{j\dagger }$, we should compare the quantum stochastic differential for 
$\hat{\phi}_{t}\left[ X\right] =J\check{\phi}_{t}\left[ X^{\dagger }\right]
J $, with the differential%
\begin{eqnarray*}
\mathrm{d}\Re \hat{p}_{t}\hat{x}_{t} &=&\Re (\mathrm{d}\hat{p}_{t}\mathrm{d}%
\hat{x}_{t}+\mathrm{d}\hat{p}_{t}\hat{x}_{t}+\hat{p}_{t}\mathrm{d}\hat{x}%
_{t})=\Re \sum_{i,j=}^{n}\kappa _{ji}\hat{\phi}_{t}[Q^{i}]\hat{c}_{t}^{j}%
\mathrm{d}t \\
&&+\Re \sum_{i=1}^{n}\hat{\phi}_{t}[Q^{i}]\hat{x}_{t}\otimes \mathrm{de}%
_{i}^{t}+\Re \hat{p}_{t}(\sum_{j=1}^{n}\hat{c}_{t}^{j}\otimes \mathrm{de}%
_{j}^{t}+\hat{g}_{t}\mathrm{d}t),
\end{eqnarray*}%
for $\Re \hat{p}_{t}\hat{x}_{t}=\hat{\phi}_{t}[X]$ obtained applying the
quantum It\^{o} formula, where we took%
\begin{equation*}
\mathrm{d}\hat{p}_{t}=\sum_{j=1}^{n}\hat{\phi}_{t}[Q^{j}]\otimes \mathrm{de}%
_{j}^{t},\;\;Q^{j}=\sum_{i=i}^{n}Q_{i}\theta ^{ij}
\end{equation*}%
for the martingale $\hat{p}_{t}=$ $\hat{\phi}_{t}\left[ I_{0}\right] $. Here 
$\mathrm{e}_{i}^{t}=\sum_{j=1}^{n}\theta _{ij}\mathrm{e}_{t}^{j}$ as we have
expressed the differential $\mathrm{d}\hat{p}_{t}=\sum_{j=1}^{n}\hat{\phi}%
_{t}[Q_{j}]\otimes \mathrm{de}_{t}^{j}$ for the martingale $\hat{p}_{t}=J%
\check{p}_{t}J$ in terms of the driving output error noises $\mathrm{e}%
_{j}^{t}$ by the real linear transformation $\boldsymbol{\theta }^{-1}$.
Comparing $\mathrm{d}\Re \hat{p}_{t}\hat{x}_{t}$ with the equation (\ref%
{eq:b25}) for $\mathrm{d}\hat{\phi}_{t}\left[ X\right] $%
\begin{equation*}
\mathrm{d}\hat{\phi}_{t}\left[ X\right] =\hat{\phi}_{t}\circ \Lambda ^{\ast
}[X](\vartheta )\mathrm{d}t+\sum_{j=1}^{n}\hat{\phi}_{t}\left[
L^{j}X+XL^{j\dagger }\right] \otimes \mathrm{de}_{j}^{t}
\end{equation*}%
written in terms of $\mathrm{e}_{j}^{t}$, we derive%
\begin{eqnarray*}
\Re \lbrack \hat{\phi}_{t}[Q^{j}]\hat{x}+\hat{p}_{t}\hat{c}_{t}^{k}] &=&\hat{%
\phi}_{t}[L^{j\dagger }X+XL^{j}]=\Re \hat{p}_{t}\hat{\rho}_{t}[L^{j\dagger
}X+XL^{j}], \\
\Re \lbrack \kappa _{ji}\hat{\phi}_{t}[Q^{i}]\hat{c}_{t}^{j}+\hat{p}_{t}\hat{%
g}_{t}] &=&\hat{\phi}_{t}[\Lambda ^{\ast }[X]]=\Re \hat{p}_{t}\hat{\rho}%
_{t}[\Lambda ^{\ast }[X]],
\end{eqnarray*}%
where $\hat{\rho}_{t}[X]=\hat{x}_{t}$. This gives (\ref{eq:b28})--(\ref%
{eq:b30}) in terms of $\Xi _{t}^{j},\;\Gamma _{t}$, defining 
\begin{equation*}
\hat{c}_{t}^{j}=\int \mathrm{Tr}_{\mathcal{H}_{0}}[\Xi _{t}^{j}[\hat{\varrho}%
_{t}](\vartheta )X(\vartheta )]\mathrm{d}\vartheta ,\;\hat{g}_{t}=\int 
\mathrm{Tr}_{\mathcal{H}_{0}}[\Gamma _{t}[\hat{\varrho}_{t}](\vartheta
)X(\vartheta )]\mathrm{d}\vartheta .
\end{equation*}%
Note that the unnormalized posterior expectation $\hat{\phi}_{t}:X\mapsto 
\hat{\phi}_{t}\left[ X\right] $, as well as its normalized version $\hat{\rho%
}_{t}\left[ X\right] =\int \mathrm{Tr}_{\mathcal{H}_{0}}[X\left( \vartheta
\right) \hat{\varrho}_{t}(\vartheta )]\mathrm{d}\vartheta $ is not CP map
but transpose-CP, in the contrast to the CP map $\check{\phi}_{t}:\mathcal{A}%
\rightarrow \mathfrak{\check{b}}_{t}$ and its normalized version $\check{\rho%
}_{t}\left[ X\right] =J\hat{\rho}\left[ X^{\dagger }\right] J$.

\textbf{Example.} Let us consider the case $n=1$ with $L_{1}=\frac{1}{2}\
Q=L_{1}^{\dagger }$, $\gamma ^{1/2}\mathrm{\check{a}}_{t}=\mathrm{a}%
_{t}^{1}=\gamma ^{-1/2}\mathrm{\hat{a}}^{t}$, where $\mathrm{a}_{t}^{1}$ is
the standard annihilation integrator $[\mathrm{a}_{s}^{1},\mathrm{a}%
_{t}^{1\dagger }]=\min \{s,t\}$ defining the input noise $\mathrm{v}%
_{t}=2\Re \mathrm{\check{a}}_{t}$ and nondemolition observation of the
commutative output process 
\begin{equation*}
\mathrm{d}Y(t)=Q(t)\mathrm{d}t+I_{0}\otimes \mathrm{de}^{t},\;\;Q(t)=U(t)^{%
\dagger }QU(t)
\end{equation*}%
by $\mathrm{e}^{t}=2\Re \mathrm{\hat{a}}^{t}=\gamma \mathrm{v}_{t}$. In this
case $\mathrm{e}_{t}=\mathrm{v}_{t}$, $y^{t}:=\mathrm{e}^{t}\delta
_{\emptyset }$ is $v_{t}^{\prime }=\gamma v_{t}$, where $v_{t}=\mathrm{v}%
_{t}\delta _{\emptyset }$ is given as $\gamma ^{-1/2}w_{t}^{1}$ by the
standard Wiener process $w_{t}^{1}$ identified with the vector process $%
\mathrm{a}_{t}^{1\dagger }\delta _{\emptyset }$ in Fock space, and the
output process $Y\left( t\right) $ on the initial state vector $\psi
_{0}\otimes \delta _{\emptyset }$ is identified with the classical output
process $y^{t}=\gamma ^{1/2}w_{t}^{1}=\gamma v_{t}$ relatively to the
probability density $p_{t}\left( v\right) \equiv \hat{p}_{t}\delta
_{\emptyset }$ with respect to the Wiener probability measure $\mathrm{P}%
_{\gamma }$ of the input Wiener process $\left\{ v_{t}\right\} $ with the
intensity $\gamma ^{-1}$. The equations (\ref{eq:b25}) and (\ref{eq:b28})
are classical stochastic equations in the linear (for the nonnormalized $%
\varphi _{t}(\vartheta ,v)\equiv \hat{\varphi}_{t}(\vartheta )\delta
_{\emptyset }$) and nonlinear (for $\varrho _{t}(\vartheta ,v)=\varphi
_{t}(\vartheta ,v)/p_{t}(v)$) posterior density operators with respect to
the output states%
\begin{equation*}
\langle \psi \otimes \delta _{\emptyset }|B_{t}\psi \otimes \delta
_{\emptyset }\rangle =\int b_{t}\left( v\right) p_{t}\left( v\right) \mathrm{%
P}_{\gamma }\left( dv\right) \equiv \langle b_{t}(\mathrm{v}),p_{t}(\mathrm{v%
})\rangle _{\emptyset }
\end{equation*}
on $B\left( t\right) =b_{t}\left( \gamma ^{-1}Y\right) $ given by any
adapted functional $b_{t}$ of $v$.

Suppose that the quantum receiver is an open oscillator (e.g. Weber's
antenna), described by the Hamiltonian $H=\frac{1}{2}\ A^{\dagger }A$, where 
$A=\mathrm{i}P+\omega Q$ and $Q,P$ are the canonical coordinate and momentum
operators: $[Q,P]=\mathrm{i}\hbar I_{0}$. Then the quantum Langevin equation
(\ref{eq:b9}) for $A(t)=U_{t}AU_{t}^{\dagger }$ is the linear one 
\begin{equation*}
\mathrm{d}A(t)+\mathrm{i}\omega A(t)\mathrm{d}t=\mathrm{i}I\otimes (\mathrm{d%
}\vartheta _{t}+\mathrm{df}_{t}),
\end{equation*}%
where $\mathrm{f}_{t}=\hbar \Im \mathrm{\check{a}}_{t}^{\dagger }$ is the
Langevin force (thermal noise) as a classical Wiener process of the
intensity $\sigma _{\gamma }^{2}=\hbar ^{2}/4\gamma $ acting on the
coordinate $Q$, defining the total force in the right hand side of the
equation as the sum $\vartheta _{t}+\mathrm{f}_{t}$ of the unknown
gravitational force $f\left( \vartheta _{t}\right) =\vartheta _{t}$ and the
thermal noise through the additive channel $\vartheta _{t}\left( \mathrm{w}%
\right) +\mathrm{f}_{t}$. In the case of Gaussian input process $\vartheta
_{t}$, corresponding to the linear $\upsilon (\vartheta )=\upsilon \vartheta 
$ and Gaussian initial state $\phi $ the optimal estimate of $X(\vartheta
_{t})=\vartheta _{t}I\equiv X\left( t\right) $ is given by the linear
posterior mean value $\hat{\vartheta}_{t}=\hat{x}_{t}$ with respect to the
output coordinate process $Y(t)$. In the standard Fock representation it is
given as the last component of the stochastic row $\hat{\boldsymbol{x}}_{t}=(%
\hat{q}_{t},\hat{p}_{t},\hat{x}_{t})$ of the posterior mean values for $%
\boldsymbol{X}(t)=(Q(t),P(t),X(t))$, satisfying the Kalman equation 
\begin{equation*}
\mathrm{d}\hat{\boldsymbol{x}}_{t}+\hat{\boldsymbol{x}}_{t}\boldsymbol{%
\Lambda }\mathrm{d}t=\boldsymbol{k}_{t}\mathrm{d\tilde{v}}_{t},\;\;%
\boldsymbol{\Lambda }=\left( 
\begin{array}{ccc}
0 & \omega ^{2} & 0 \\ 
-1 & 0 & 0 \\ 
0 & -\upsilon & \upsilon%
\end{array}%
\right) ,
\end{equation*}%
where $\boldsymbol{k}_{t}=(k_{t}^{11},k_{t}^{12},k_{t}^{13})$ is the first
row of the symmetric $3\times 3$--matrix $\boldsymbol{K}_{t}^{\mathrm{s}%
}=(k_{t}^{ij})$ satisfying the Riccati equation 
\begin{equation*}
\frac{\mathrm{d}}{\mathrm{d}t}\ \boldsymbol{K}_{t}+\boldsymbol{K}_{t}%
\boldsymbol{\Lambda }+\boldsymbol{\Lambda }^{\top }\boldsymbol{K}_{t}+\frac{1%
}{\gamma }\boldsymbol{k}_{t}^{\top }\boldsymbol{k}_{t}=\boldsymbol{\Upsilon }%
,\;\;\boldsymbol{\Upsilon }=\left( 
\begin{array}{ccc}
0 & 0 & 0 \\ 
0 & \sigma ^{2}+\sigma _{\gamma }^{2} & \sigma ^{2} \\ 
0 & \sigma ^{2} & \sigma ^{2}%
\end{array}%
\right) ,
\end{equation*}%
with an initial symmetric covariance matrix $\boldsymbol{K}_{0}^{\mathrm{s}}$
of $\left( Q,P,\vartheta _{0}\right) $ and $\mathrm{\tilde{v}}_{t}=\mathrm{v}%
_{t}-\gamma ^{-1}\hat{q}_{t}$. The pair $(\hat{\boldsymbol{x}}_{t},%
\boldsymbol{K}_{t})$ defines the posterior (normalized) Gaussian state of
the quantum system with input signal $x_{t}=\vartheta _{t}$ and the mean
square error $\langle \overset{\sim }{\vartheta }_{t}^{2}\rangle $, $\overset%
{\sim }{\vartheta }_{t}=\vartheta _{t}-\hat{\vartheta}_{t}$ is given by the
component $k_{t}^{33}$ of the posterior correlation matrix $\boldsymbol{K}%
_{t}$.

A posterior dynamics of the quantum system under another nondemolition
measurement of the received electromagnetic field by a photon counter is
considered in \cite{bib:b14}.


\end{document}